# An Effective Approach for Web Document Classification using the Concept of Association Analysis of Data Mining

Rajendra Kumar Roul
BITS, Pilani - K.K. Birla, Goa Campus
Zuarinagar, Goa - 403726, India
(rkroul@goa.bits-pilani.ac.in)

S.K Sahay
BITS, Pilani - K.K. Birla, Goa Campus
Zuarinagar, Goa - 403726, India
(ssahay@goa.bits-pilani.ac.in)

*Abstract*- **Exponential growth of the web increased the importance of web document classification and data mining. To get the exact information, in the form of knowing what classes a web document belongs to, is expensive. Automatic classification of web document is of great use to search engines which provides this information at a low cost. In this paper, we propose an approach for classifying the web document using the frequent item word sets generated by the Frequent Pattern (FP) Growth which is an association analysis technique of data mining. These set of associated words act as feature set. The final classification obtained after Naïve Bayes classifier used on the feature set. For the experimental work, we use Gensim package, as it is simple and robust. Results show that our approach can be effectively classifying the web document.**

## I. Introduction

Web document classification is the process of classifying documents into predefined categories based on their content. The classifiers used for this purpose should be trained from the web documents that are already classified. The task is to assign a document to one or more classes or categories. This may be done "manually" (or "intellectually") or algorithmically. Manual classification cost more. The intellectual classification of documents has mostly been the province of library science, while the algorithmic classification of documents is used mainly in information science and computer science. The problems are overlapping; however there is also interdisciplinary research on documents classification. The documents to be classified may be texts, images, music, etc. Each kind of document possesses its special classification problems. Documents may be classified according to their subjects or according to other attributes. Web document classification is the primary requirement for search engines, which retrieve documents in response to the user query. Documents classification or text categorization (as used in information retrieval context) is the process of assigning a document to a predefined set of categories based on the document content. Documents classification can be applied as an information filtering tool and can also be used to improve the retrieval results from a query process. Classification is one of the main data analysis techniques and deals with the categorizing a new data entry into one of the categories based on the values of different attributes. In general, classification algorithm needs to train a model based on pre-classified training documents. Once the model is ready, we can subject the test documents for evaluation through that model and that brings the classification process to an end.

In this paper, we proposed an approach for automatically classifying web documents into a set of categories using FP-growth and Naïve Bayes techniques. In our approach, we have given a set of example documents. We preprocess the documents by parsing and removing the stop words, doing stemming [10] and extracted noun as keywords. Then we apply FP-growth method [9] to find the frequent item word sets from each document.  The documents are treated as transactions and the set of frequently occurring words are viewed as a set of items in the transaction. The new documents are classified by applying Naïve Bayes technique on these derived features sets.

The paper is organized as follows: Section 2 covers the related work based on different classification techniques used for web document. Section 3 describes the materials and methods used in our proposed approach. In section 4, we describe the proposed approach adopted for classification. The results and discussions are covered in section 5 and finally conclusion is presented in section 6.

## II. Related Work

Web document classification has been widely studied in the past few years. Much research work has been done in this area. Chakrabarti et al.[1] used predicted labels of neighboring documents to reinforce classification decisions for a given document. Qi and Davison [2] summarizes the various concepts used for automatic web page classification with respect to recent works. A dynamic and hierarchical classification system that is capable of adding new categories as required, organizing the web pages into a tree structure, and classifying web pages by searching through only one path of the tree structure is proposed in [3]. The test results show that the proposed single-path search technique reduces the search complexity and increases the accuracy by 6% comparing to related algorithms. Positive Example Based Learning (PEBL) [4] is a framework for Web page classification which eliminates the need for manually collecting negative training examples in preprocessing. Oh et al. [5] proposed a practical method for exploiting hypertext structure and hyperlink information. They modified the Naive Bayes algorithm to classify documents by using neighboring documents that were similar to the target document. Both the predicted labels and the text contents of the neighboring documents were used to assistant classification. The experimental results on an encyclopedia corpus that contains hyperlinks validate their algorithms. F¨urnkranz [6] also reported a significant improvement in classification accuracy when using the link-based method as opposed to the full-text alone on 1,050 pages of the Web KB corpus, although adding the entire text of "neighbor documents" seemed to harm the ability to classify pages [1]. A. Sun et al. [7] claimed that the combination of the plain text, the anchor text and the title can get a large improvement on F1 measure compared with full-text. Liu et al. [8] present an Entity-Based web page classification algorithm, which can be embedded in search engines easily. In this algorithm, an Entity System is built to classify web pages immediately before indexing jobs. Our work is an effective ways to classify the web documents by using FP-Growth and Naïve Bayes techniques.

## III. Materials and Methods

Vector Space Model

In vector space model, each document is defined as a multidimensional vector of keywords in Euclidean space whose axis correspond to the keyword i.e., each dimension corresponds to a separate keyword [11]. The keywords are extracted from the document and weight associated with each keyword determines the importance of the keyword in the document. Thus, a document is represented as, $D_j = (w_{1j}, w_{2j}, w_{3j}, w_{4j}, \ldots\ldots\ldots, w_{nj})$ where, $w_{ij}$ is the weight of term i in document j indicating the relevance and importance of the keyword words.

TF-IDF

TF is the measure of how often a word appears in a document and IDF is the measure of the rarity of a word within the search index. Combining TF-IDF[11] is used to measure the statistical strength of the given word in reference to the query. Mathematically, $TF_i = n_i/(\Sigma k n_k)$ where, $n_i$ is the number of occurrences of the considered terms and $n_k$ is the number of occurrences of all terms in the given document. $IDF_i = (\log N)/df_i$ where, N is the number of occurrences of the considered terms and $df_i$ is the number of documents that contain term i.
$$TF\text{-}IDF = TF_i \times IDF_i$$

Gensim

Gensim package[12] is a python library for vector space modeling, aims to process raw, unstructured digital texts ("plain text"). It can automatically extract semantic topics from documents, used basically for the Natural Language Processing (NLP) community. Its memory (RAM) independent feature with respect to the corpus size allows to process large web based corpora.

*A.* FP-growth Algorithm:

Algorithm 1 (FP-tree construction)**:**

Input: A transaction database *D* and a minimum support threshold *ξ*.

Output: FP-tree, the frequent-pattern tree of *D*.

Method: The FP-tree is constructed as follows.

1. Collect the set of frequent items($F_{items}$) and their support counts after scanning the transaction database(D) once. Sort $F_{items}$ according to descending support count as $L_{freq}$, the list of frequent items.

2. Create the root of an FP-tree, and label it as "null". For each transaction $I_{trans}$ in D do the following,

Select and sort the frequent items in $I_{trans}$ according to the order of $L_{freq}$. Let the sorted frequent list in $I_{trans}$ be [*e* | $E_{list}$], where *e* is the first element and $E_{list}$ is the remaining list. Call *insert_tree*([*e* | $E_{list}$], *T*), which is performed as follows:

Procedure *insert_tree(* [*e* | $E_{list}$], *T*)

if *T* has a child *N* such that *N*.item-name=*e*.item-name, then increment *N*'s count by 1; else create a new node *N*, and let its count be 1, its parent link be linked to *T*, and its node-link to the nodes with the same item-name via the node-link structure. If $E_{list}$ is nonempty, call *insert_tree*($E_{list}$, *N*) recursively.

Algorithm 2 (FP-growth: *Mining frequent patterns with FP-tree by pattern fragment growth*).

Input: A database *D*, represented by FP-tree constructed according to Algorithm 1, and a minimum support threshold $\xi$.

Output: The complete set of frequent patterns.

Method: *call FP-growth*(FP-tree, *null*).

Procedure *FP-growth*(*Tree, α*) {

*if Tree* contains a single prefix path *then* {

*let P* be the single prefix-path part of *Tree*;

*let Q* be the multipath part with the top branching node replaced by a *null* root;

*for each* combination (denoted as *β*) of the nodes in the path *P* do generate pattern *β* ∪ *α* with *support = minimum support of nodes in β*;

*let freq_ pattern_ set*(*P*) be the set of patterns so generated; *}*

*else let Q* be *Tree*;

*for each* item $a_i$ in *Q* do {

generate pattern *β* = $a_i$ ∪ *α* with *support* = $a_i$.*support*;

construct *β*'s conditional pattern-base and then *β*'s conditional FP-tree $Tree_β$ ;

*if* $Tree_β$ ≠ Φ

*then call FP-growth*($Tree_β$, *β*);

*let freq_ pattern_ set*(*Q*) be the set of patterns so generated; *}*

*return*(*freq_ pattern_ set*(*P*) ∪ *freq_ pattern_ set*(*Q*) ∪ (*freq_ pattern_ set*(*P*)×*freq_ pattern_ set*(*Q*)))

*}*

*B*. Naïve Bayes Technique:

The naïve-bayes classifier for probability calculation is defined as follows:

$$Y_{NB} = P(Y_j) \Pi P(a_i | Y_j) \quad (Eq.1)$$

Where, $Y_j$ represents the class j. $Y_{NB}$ is the probability that a document having $a_i$ attribute belongs to class j. The calculation for first term is based on the fraction of each target class in the training data. Then the second term of the equation is calculated by the following equation after adopting m-estimate approach in order to avoid zero probability value,

$$\frac{n_k + 1}{n + vocabulary} \quad (Eq.2)$$

Where, n = Total no of frequent word sets in all training examples whose class value is j

$n_k$ = No. of times the frequent word set found among all the training examples whose target class value is j.

vocabulary = the total number of distinct word sets found within all the training data.

## IV. The Proposed Approach

1. *Preprocessing Training Data Set* (*TRD*) *and Test Data Set* (*TED*):
   - Remove the stop and unwanted words from both *TRD* and *TED*.
   - Select noun as the keywords from both data set and remove duplicate keywords from each document.
   - Do stemming using porter algorithm on both data sets.
   - Save each processed *n* pages of *TRD* as document $D_k$, where k = 1, 2, 3,…, *n* and each processed *m* pages of *TED* as document $TED_j$, where j = 1, 2, 3,…, *m*.

2. *Create term document matrix:*
   Term document matrix, *T*, is created by counting the number of occurrences of each term in each document $D_k$. Each row $t_i$ of *T* shows a term's occurrence in each document $D_k$.

3. *Extraction of frequent sets:*

   FP-growth algorithm is used to generate frequent word sets from the term document matrix *T* using the value of minimum support, *min_sup*, given as an input and stored in *F*. Calculate the probability values of each frequent word sets stored in *F* using Naïve Bayes method.

4. *Finding matching word set(s):*

   Search for matching word set(s) or its subset (containing items more than one) in the list of word sets collected from *F* with that of subset(s) of word of new test document using regular expression search.

5. *Calculate the probability values of target class:*

   a. Collect the corresponding probability values of matched word set(s) for each target class

   b. Calculate the probability values for each target class from naïve based classification approach using Eq. 1 and Eq. 2.

6. *Assignment the new document to that target class which has highest probability values.*

*Flow Diagram*: The following figure shows the steps to obtain the classification of web documents which has discussed in the proposed approach.

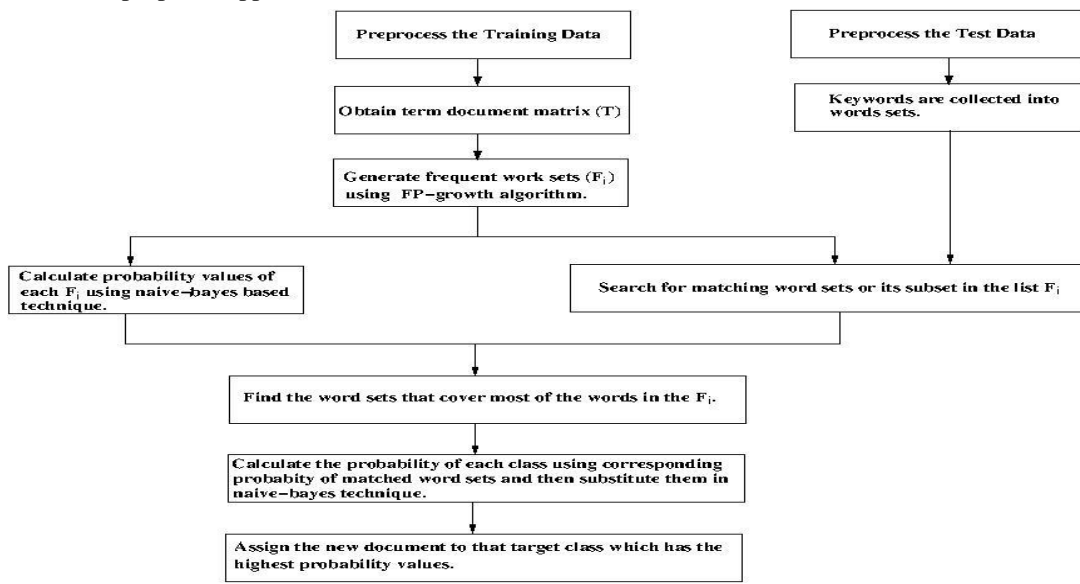

Fig.1: Classification of web document using FP-growth and Naïve Bayes techniques.

V. Results and Discussions

To illustrate our approach, we took ten training documents which shown in table 1. These documents broadly categories into two topics namely *Social Network* (D1 to D6) and *Computer Network* (D7 to D10). We have taken documents as transactions. The minimum support taken as *two*. Table 2 shows the preprocessing training documents. The term document matrix of the preprocessed training documents shown in table 3. After using FP-growth algorithm on term-document matrix of training set, we found frequent word set and their number of occurrence in each predefined classes called *Social Network* and *Computer Network*. The FP-tree for both *Social Network* and *Computer Network* shown in Fig.2 and Fig.3 respectively. The probability values of each word set found out by using naïve bayes method which has been shown in table 4. We tested the trained classifier through a test set of five documents shown in table 5 and their preprocessing data in table 6. Table 7 shows the classification of test documents into their corresponding target class. The data collected from FP tree of Fig.2 and Fig.3 used to construct the table 4 and table 7 shown below. The details of these tables are shown in appendix.

Total number of word set = 24
Total number word set from *Social Network* = 13
Total number of word set from *Computer Network* = 11
Prior probability for *Social Network* = (6/10) = 0.6
Prior probability for *Computer Network* = (4/10) = 0.4
One example each for table 4 and table 7 has been discussed below:
Table 4: P(website, people | SN) = (3+1) / (13+24) = 0.108 and P(website, people | CN) = (0+1) /(11+24) =0.029

Table 7 for T1: The word set that covers most of the words in the document is {network, computer, data}

P (SN) = 0.6, P (CN) =0.4, P (network, computer, data |SN) = 0.027, P (network, computer, data |CN) = 0.086.

Probability of document belonging to *Social Network* = 0.6 x 0.027 = 0.0162.

Probability of document belonging to *Computer Network* = 0.4 x 0.086 = 0.0344.

Therefore the test document T1 belongs to class *Computer Network*.

## VI. Conclusion

We used FP-growth technique along with Naïve Bayes to classify the web document. The purpose of FP-growth is to generate frequent word sets from training data set. We treated documents as transactions and the set of frequently occurring words as a set of items in the transaction. The new documents are classified by applying Naïve Bayes method on these derived sets. The experimental results show that our approach is an effective way to classify the web document which will be used by search engines. This work can be further extended by considering the semantic nature of the web document instead of keyword based.

Appendix

| TRD_id | Documents |
|---|---|
| D1 | A dedicated website or other application that enables people to communicate with each other by posting information. |
| D2 | Uses special sites to allow people to create a profile and form communities based on common interests. |
| D3 | A social network service is an online service, platform, or site that focuses on building and reflecting of social networks or social relations among people. |
| D4 | Social networks can be thought of as communities based upon interest or commonality that use the internet to connect the people of the network. |
| D5 | A group of people who exchange information and experience for professional or social purposes. |
| D6 | Networking is establishing informal communities of contacts among people with common social and business interests as a source of prospects, for the exchange of information, and for support. |
| D7 | A computer network is a group of computers and devices interconnected by communications channels that facilitate communications among users and allow users to share resources and data. |
| D8 | Computer networking is the joining of two or more computers in order for them to communicate or jointly access a server. |
| D9 | A group of two or more computers linked by cables or wireless signals or both, which can communicate with one another using network protocols. |
| D10 | A group of computers together with the sub-network or inter-network through which they can exchange data is called a computer network. |

Table 1: Training Documents.

| TRD_id | Keywords |
|---|---|
| D1 | website application people information |
| D2 | website people profile community interest |
| D3 | network service  platform website relation people |
| D4 | network community interest commonality internet people |
| D5 | group people information experience purpose |
| D6 | network contact community people interest prospect information support |
| D7 | computer network group device channel communication user resource data |
| D8 | computer network server |
| D9 | group computer cable signal network protocol |
| D10 | group computer network  data |

Table 2: Training Documents after preprocessing.

| Keywords | D1 | D2 | D3 | D4 | D5 | D6 | D7 | D8 | D9 | D10 |
|---|---|---|---|---|---|---|---|---|---|---|
| website | 1 | 1 | 1 | 0 | 0 | 0 | 0 | 0 | 0 | 0 |
| application | 1 | 0 | 0 | 0 | 0 | 0 | 0 | 0 | 0 | 0 |
| people | 1 | 1 | 1 | 1 | 1 | 1 | 0 | 0 | 0 | 0 |
| information | 1 | 0 | 0 | 0 | 1 | 1 | 0 | 0 | 0 | 0 |
| profile | 0 | 1 | 0 | 0 | 0 | 0 | 0 | 0 | 0 | 0 |
| community | 0 | 1 | 0 | 1 | 0 | 1 | 0 | 0 | 0 | 0 |
| interest | 0 | 1 | 0 | 1 | 0 | 1 | 0 | 0 | 0 | 0 |
| network | 0 | 0 | 1 | 1 | 0 | 1 | 1 | 1 | 1 | 1 |
| service | 0 | 0 | 1 | 0 | 0 | 0 | 0 | 0 | 0 | 0 |
| platform | 0 | 0 | 1 | 0 | 0 | 0 | 0 | 0 | 0 | 0 |
| relation | 0 | 0 | 1 | 0 | 0 | 0 | 0 | 0 | 0 | 0 |
| commonality | 0 | 0 | 0 | 1 | 0 | 0 | 0 | 0 | 0 | 0 |
| internet | 0 | 0 | 0 | 1 | 0 | 0 | 0 | 0 | 0 | 0 |
| group | 0 | 0 | 0 | 0 | 1 | 0 | 1 | 0 | 1 | 1 |
| contact | 0 | 0 | 0 | 0 | 0 | 1 | 0 | 0 | 0 | 0 |
| experience | 0 | 0 | 0 | 0 | 1 | 0 | 0 | 0 | 0 | 0 |
| purpose | 0 | 0 | 0 | 0 | 1 | 0 | 0 | 0 | 0 | 0 |
| prospect | 0 | 0 | 0 | 0 | 0 | 1 | 0 | 0 | 0 | 0 |
| support | 0 | 0 | 0 | 0 | 0 | 1 | 0 | 0 | 0 | 0 |
| computer | 0 | 0 | 0 | 0 | 0 | 0 | 1 | 1 | 1 | 1 |
| device | 0 | 0 | 0 | 0 | 0 | 0 | 1 | 0 | 0 | 0 |
| channel | 0 | 0 | 0 | 0 | 0 | 0 | 1 | 0 | 0 | 0 |
| communication | 0 | 0 | 0 | 0 | 0 | 0 | 1 | 0 | 0 | 0 |
| user | 0 | 0 | 0 | 0 | 0 | 0 | 1 | 0 | 0 | 0 |
| resource | 0 | 0 | 0 | 0 | 0 | 0 | 1 | 0 | 0 | 0 |
| data | 0 | 0 | 0 | 0 | 0 | 0 | 1 | 0 | 0 | 1 |
| server | 0 | 0 | 0 | 0 | 0 | 0 | 0 | 1 | 0 | 0 |
| cable | 0 | 0 | 0 | 0 | 0 | 0 | 0 | 0 | 1 | 0 |
| signal | 0 | 0 | 0 | 0 | 0 | 0 | 0 | 0 | 1 | 0 |
| protocol | 0 | 0 | 0 | 0 | 0 | 0 | 0 | 0 | 1 | 0 |

Table 3: Term Document matrix.

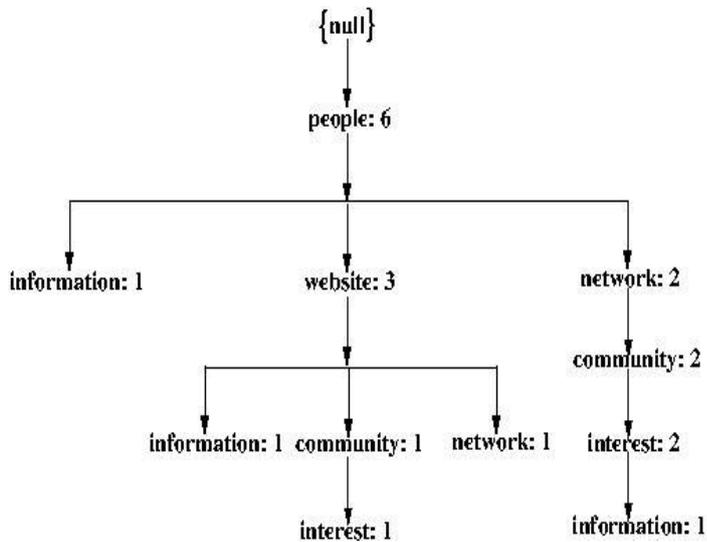

Fig. 2: FP Tree (*Computer Network*)

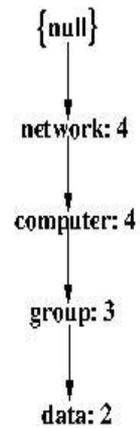

Fig. 3: FP Tree (*Social Network*)

| S.no | Word sets | *Social Network* | *Computer Network* |
|---|---|---|---|
| 1 | people, website | 0.108 | 0.029 |
| 2 | people, information | 0.108 | 0.029 |
| 3 | people, community | 0.108 | 0.029 |
| 4 | people, interest | 0.108 | 0.029 |
| 5 | people, network | 0.108 | 0.029 |
| 6 | network, interest | 0.081 | 0.029 |
| 7 | network, community | 0.081 | 0.029 |
| 8 | community, interest | 0.108 | 0.029 |
| 9 | network, computer | 0.027 | 0.143 |
| 10 | network, group | 0.027 | 0.114 |
| 11 | network, data | 0.027 | 0.086 |
| 12 | computer, group | 0.027 | 0.114 |
| 13 | computer, data | 0.027 | 0.086 |
| 14 | group, data | 0.027 | 0.086 |
| 15 | people, community, interest | 0.108 | 0.029 |
| 16 | people, network, community | 0.081 | 0.029 |
| 17 | people, network, interest | 0.081 | 0.029 |
| 18 | network, community, interest | 0.081 | 0.029 |
| 19 | network, computer, group | 0.027 | 0.114 |
| 20 | network, computer, data | 0.027 | 0.086 |
| 21 | network, group, data | 0.027 | 0.086 |
| 22 | computer, group, data | 0.027 | 0.086 |
| 23 | network, computer, group, data | 0.027 | 0.086 |
| 24 | people, network, community, interest | 0.081 | 0.029 |

Table 4: Probability values of word sets.

| TED_Id | Keywords |
|---|---|
| T1 | In a network environment, authorized users may access data and information stored on other computers on the network. |
| T2 | These communities of hypertexts allow for the sharing of information, ideas and interests among people, an old concept placed in a digital environment via websites. |
| T3 | A network firewall guards a group of computer against unauthorized data access. |
| T4 | A different community of people shares their common interest on websites. |
| T5 | A community of interest is a network of people assembled around a topic of common information. |

Table 5: Test Documents.

| TED_Id | Keywords |
|---|---|
| T1 | network, authorize, environment, user, access, data, information, computer |
| T2 | community, hypertext, share, information, idea, interest, people, concept, digital, environment, website |
| T3 | network, group, firewall, guard, computer, unauthorize, data, access |
| T4 | different, community, people, share, common, interest, website |
| T5 | community, interest, network, people, assemble, around, topic, common, information |

Table 6: Test Documents after preprocessing.

| TED_Id | Word sets | Prob. of $T_i$ belongs to the class *Social Network* | Prob. of $T_i$ belongs to the class Computer Network | Target class(max probability value) |
|---|---|---|---|---|
| T1 | {network, computer, data} | 0.0162 | 0.0344 | *Computer Network* |
| T2 | {people, community, interest} {people, information} {people, website} | 0.0007558 | 0.0000098 | *Social Network* |
| T3 | {network, group, computer, data} | 0.0162 | 0.0344 | *Computer Network* |
| T4 | {people, community, interest} {people, website} | 0.0069984 | 0.0003364 | *Social Network* |
| T5 | {people, network, community, interest} {people, information} | 0.0052488 | 0.0003364 | *Social Network* |

Table 7: Test Documents and their respective target classes.